\documentclass[sigconf]{acmart}

\pdfoutput=1
\usepackage{amsmath,amssymb,amsfonts}
\usepackage{algorithmic}
\usepackage{graphicx}
\usepackage{textcomp}
\usepackage{xcolor}
\usepackage{url}
\usepackage{ulem}
\usepackage{hyperref}
\usepackage{listings}
\usepackage{subfigure} 
\usepackage{multirow}
\usepackage{pifont}
\usepackage{framed}
\usepackage{threeparttable}
\usepackage[linesnumbered,ruled,vlined]{algorithm2e}
\DeclareMathOperator*{\argmax}{argmax}
\lstdefinestyle{oc_code}{
	language=c,
	breaklines=true,
	frame=lines, 
	basicstyle=\footnotesize,
	numberstyle=\footnotesize,
	language=[Objective]C,
}
\usepackage{enumitem}

\usepackage{breakurl}           
\usepackage{url}                
\usepackage{xcolor}             
\usepackage[]{hyperref}         
\hypersetup{                    
	colorlinks,
	linkcolor={green!80!black},
	citecolor={red!70!black},
	urlcolor={blue!70!black}
}
\newcommand {\tool} {{\textsc{iLibScope}}\xspace}

\newcommand{\code}[1]{{\fontfamily{cmtt}\fontseries{m}\fontshape{n}\selectfont\small{#1}}}
\newcommand{\tab}{\hspace*{1em}}

\AtBeginDocument{%
  \providecommand\BibTeX{{%
    \normalfont B\kern-0.5em{\scshape i\kern-0.25em b}\kern-0.8em\TeX}}}

\renewcommand\footnotetextcopyrightpermission[1]{} %

\begin{document}

\title{\tool: Reliable Third-Party Library Detection for iOS Mobile Apps}

\author{Jingyi Guo}
\email{allblues020201@gmail.com}
\affiliation{%
}
\author{Min Zheng}
\affiliation{%
 \institution{Alibaba}
}
\author{Yajin Zhou}
\affiliation{%
 \institution{Zhejiang University}
}
\author{Haoyu Wang}
\affiliation{%
 \institution{Huazhong University of Science and Technology}
}
\author{Lei Wu}
\affiliation{%
 \institution{Zhejiang University}
}
\author{Xiapu Luo}
\affiliation{%
 \institution{The Hong Kong Polytechnic University}
}
\author{Kui Ren}
\affiliation{%
 \institution{Zhejiang University}
}
\renewcommand{\shortauthors}{J. Guo, M. Zheng, Y. Zhou, H. Wang, L. Wu, X. Luo, K. Ren}

\begin{abstract}

Vetting the security impacts introduced by third-party libraries in iOS mobile apps requires a reliable library detection technique. 
Especially when a new vulnerability (or a privacy-invasive behavior) was discovered in a third-party library, 
there is a practical need to precisely identify the existence of third-party libraries and their versions for iOS apps. 
However, few studies have been proposed to tackle this problem, and they all suffer from the code duplication problem in different libraries.

In this paper, we focus on third-party library detection in iOS mobile apps. 
Given an app, we aim to identify the integrated libraries and pinpoint their versions (or the version range).
To this end, we first conduct an in-depth study on iOS third-party libraries to reveal the relationship between them to demystify code duplication. By doing so, we have the following two key observations: 1) even though two libraries can share some classes, however, the shared classes
cannot be integrated into an app at the same time without causing a class name conflict; and 2) code duplication between multiple versions of two libraries can vary.
Based on these observations, we propose a novel profile-based similarity comparison approach to perform the detection.
Specifically, we build a library database consists of original library binaries with distinct versions. 
After extracting profiles for each library version and the target app, we conduct a similarity comparison to find the best matches.

We implemented this approach in \tool{}. 
We also built a benchmark consists of 5,807 apps with 10,495 library integrations and applied our tool to it. 
Our evaluation shows that \tool{} achieves a recall exceeds 99\% and a precision exceeds 97\%.
We also applied \tool{} to detect the presence of well-known vulnerable third-party libraries
in real-world iOS mobile apps to show the promising usage of our tool.
It successfully identified $405$ vulnerable library usage from $4,249$ apps. 

\end{abstract}


\maketitle

\section{Introduction}

Third-party libraries play an important role in iOS application (app in short)
ecosystem. They facilitate the development process by providing commonly useful functions and
additional services like app analytics or app monetization (e.g., ad libraries).
A previous study~\cite{orikogbo2016crios} aggregated a dataset of $43,404$ iOS apps and
found that third-party libraries (developed by Objective-C) have occupied over $60\%$ of
classes in iOS apps on average.

Such a large amount of third-party libraries in apps can
introduce security issues~\cite{CodeQual0:online}. 
First, they may leak the user's private data on behalf of the host app that uses them.
For instance, Egele et al.~\cite{egele2011pios} found $55\%$ of tested apps
include either advertisement or tracking libraries.
All these libraries transmitted the unique device ID to third parties. 
Second, in a cross-platform analysis of potentially harmful libraries, Chen et al.~\cite{chen2016following}
discovered that most Android-side harmful behaviors were preserved on their iOS counterparts,
such as reading from the keychain, stealthily recording audio and video, and attempting to make phone calls. 
Third, the vulnerability existing in libraries can bring risks to host apps.
Some popular libraries have been reported to have vulnerabilities, like AFNetworking~\cite{CVECVE2081:online}, SSZipArchive~\cite{Releasev85:online}, and  GCDWebServer~\cite{CVECVE2060:online}. 
Besides, in a recent work, Tang et al.~\cite{tangios} found
that the use of certain third-party libraries listening for remote connections is a common source of
vulnerable network services. 

Obviously, vetting the security impacts introduced by third-party libraries requires 
a reliable library detection technique for iOS apps. Especially when a new
vulnerability (or a privacy-invasive behavior) was detected in a third-party
library, we need a tool to quickly scan the app to report whether the vulnerable (or aggressive) version of the library is in apps. 
Namely, this requires \textit{a generic solution to precisely identify the existence of third-party libraries and their versions}~\footnote{Or the range of possible versions if the exact versions cannot be determined.} for iOS apps.

Unfortunately, few works have been proposed to tackle this problem, while previous systems cannot satisfy this requirement.
Early work~\cite{egele2011pios,han2014android} applies a whitelist-based method to identify third-party classes in apps, 
and the subsequent studies~\cite{orikogbo2016crios, chen2016following} resort to code dependency to recover library instances from apps. 
Recently, Tang et al.~\cite{tangios} identify third-party
libraries from apps based on a call stack similarity analysis. But
only a small portion of network service libraries are focused.
However, based on our investigation (see Section~\ref{sec:motivating}), these systems all suffer from a technical challenge, i.e., \textit{code duplication in different libraries}.
Due to the complex relationships between libraries, such as a dependency relationship or sharing code in a workgroup, libraries can have duplicated code.
On the one hand, matching an app against original libraries can lead to false positives due to the duplicated region of distinct libraries.
On the other hand, a library instance recovered by code dependency may include classes from more than one library due to the inter-library relationships.


Alternatively, though the library detection for the Android platform has been well explored in recent years~\cite{grace2012unsafe,book2013longitudinal,narayanan2014addetect,crussell2014andarwin,chen2014achieving,liu2015efficient,wang2015wukong,ma2016libradar,backes2016reliable,li2017libd,glanz2017codematch,zhang2018detecting,wang2018orlis,zhang2019libid},
the proposed methods cannot be applied to iOS.
Most of them build their methods based on the Java package structure and code dependency, which are used to identify (or recover) library candidates from apps. 
While for iOS apps, first, app code is organized in a flat space, thus there is no structural information to help group code into potential library candidates. 
Second, the semantic gaps between the Dalvik VM bytecode (Android app) and the native ARM binary (iOS app)
make it challenging to apply the methods that work for Android apps on iOS apps.

\smallskip
\noindent\textbf{Our approach}\tab
In this paper, we propose a novel profile-based similarity comparison approach to perform the detection.
Specifically, we first construct a library database consists of libraries of different versions. 
Then, for each version of the library, we build \textit{profiles} (or signatures) for the classes and methods inside the library. 
Here two types of profiles, i.e., the class-level profile and the method-level (code-level) profile, are extracted from a given library.
As a result, after building the profiles for the target app, we can conduct a similarity comparison to determine the best matches.
Note that, we perform the library detection and a coarse-grained version identification based on
class-level profiles only, and a fine-grained version pinpointing based on code-level profiles.

To overcome the code duplication issue, we propose our solution based on two observations (see Section~\ref{sec:motivating}).
First, even though two libraries can share some classes, however, the shared classes
cannot be integrated into an app at the same time without causing a class name conflict.
Second, code duplication between multiple versions of two libraries can vary. For instance,
a particular version of library A that shares code with a version of library B, does not necessarily share code with another version of library B.

Accordingly, if some classes inside the app once were seen in both $A$ and $B$, 
we need to determine which library introduces them exactly. 
To this end, we try to find a version pair $(x, y)$ that $A_x$ and $B_y$ have no common classes so that they can be integrated into the app at the same time.
To make this idea a general solution, we propose an algorithm to identify the provenance of such duplicated code when we find them inside the app.

We implemented this approach in \tool{}, a reliable and scalable third-party library detection tool for iOS apps with the capacity of pinpointing library versions. 
We build a database consists of profiles for $5,768$ library versions from $319$ distinct libraries, and a benchmark consists of apps with known version information.
According to our evaluation, our two-level profile has a good uniqueness. 87.9\% of our class-level profiles can limit the candidate versions to 5 ($\le 5$), and 96.5\% of code-level profiles can achieve the same result.
We applied our tool to 5,807 apps with 10,495 libraries. Our evaluation shows that 
it achieves a recall exceeds 99\% and a precision exceeds 97\%.
We also applied \tool to detect the presence of vulnerable third-party libraries
in real-world iOS mobile apps to show the usage of our system.
It successfully identified $405$ vulnerable library usage from $4,249$ apps. 

\vspace{-0.1in}

\smallskip
\noindent
\textbf{Contributions}
In summary, this paper makes the following main contributions:
\vspace{-0.1in}
\begin{itemize}
	\item We have studied the code duplication issues in iOS third-party libraries,
	which obstacles the library detection. We performed an empirical study on a dataset of
	of $319$ distinct libraries with $5,768$ versions and reported our observations.
	These observations guide the design of our solution.
	
	\item To the best of our knowledge, \tool is the first effective approach of
	third-party library detection in iOS apps with the capability of identifying
	library versions (or version range). 

	\item The evaluation result suggested that \tool can
	achieve a recall exceeds 99\% and a precision exceeds 97\%.
	We have also applied \tool to detect the presence of vulnerable
	third-party libraries in
	real-world iOS mobile apps, with a result of identifying $405$ vulnerable library usage
	from $4,249$ iOS apps.

\end{itemize}

\section{Background and technique challenge}
\label{sec:motivating}

In this section, we first summarize different ways to integrate third-party libraries.
After that, we reveal the relationship between libraries to demystify \textit{code duplication} based on our collected libraries, and measure the impact of code duplication.
Finally, we present our solution accordingly.

Although two main languages are used to develop iOS apps (i.e., Objective-C and Swift), in this paper, we focus on apps and libraries written in Objective-C.
Objective-C has a longer history than Swift, which results in a more abundant third-party library repository for study.
Besides, we choose the binary files of apps and libraries as targets to research considering that the source code is not always available.
We automatically compile a library released in source code into binaries during the collection process (see Section~\ref{subsubsec:libdb}).



\subsection{Library integration}
\label{sec: integrating method}
For app developers, there are mainly two ways to integrate third-party libraries into their apps.
The first is via third-party library management tools such as CocoaPods~\cite{CocoaPods} and Carthage~\cite{Carthage}, and the second is copy-paste.
When there is a heavy requirement for code customization, copy-paste manner is of great flexibility; 
otherwise, integrating libraries by tools is much more convenient, especially if there are complex dependency requirements.

Meanwhile, this rule also applies to library developers.
However, different from apps (i.e., the library code will be integrated into the app regardless of the manners), 
adopting different integration methods can result in distinct library composition.
For example, when library $A$ includes $B$ through copy-paste, $A$ takes over this $B$ copy and is shipped with it during distribution.
In contrast, integrating $C$ through management tools results in a loose relationship.
First of all, $A$ does not need to include $C$ into its repository. $C$ will be integrated into the app automatically during the integration of $A$.
Second, $A$ can use flexible strategies to integrate a dynamic copy of $C$, for instance, the newest patched version.
It is important to note that we do not take a loosely integrated $C$ as part of $A$ during the library collection as well as the profile extraction process, considering the space consumption and the profile (signature) stability.
While for the copy-paste inclusion, in most cases, the integration is obscure unless we conduct a library detection on libraries themselves.
Besides, the pasted code can be customized.
Therefore, we take the tightly integrated $B$ copy as part of $A$.

\subsection{Code duplication}

To understand code duplication between libraries, we first conduct an empirical study to characterize the relationship between libraries by investigating collected libraries~\footnote{Totally 319 unique libraries with 5,768 versions, see Section~\ref{subsubsec:libdb}}.
Then, we show the statistics of code duplication among our library database.
At last, we present our solution based on a case study.

\subsubsection{Inter-library relationships}
\label{sec: inter-library relationships}

\begin{figure}
\centerline{\includegraphics[scale=0.22]{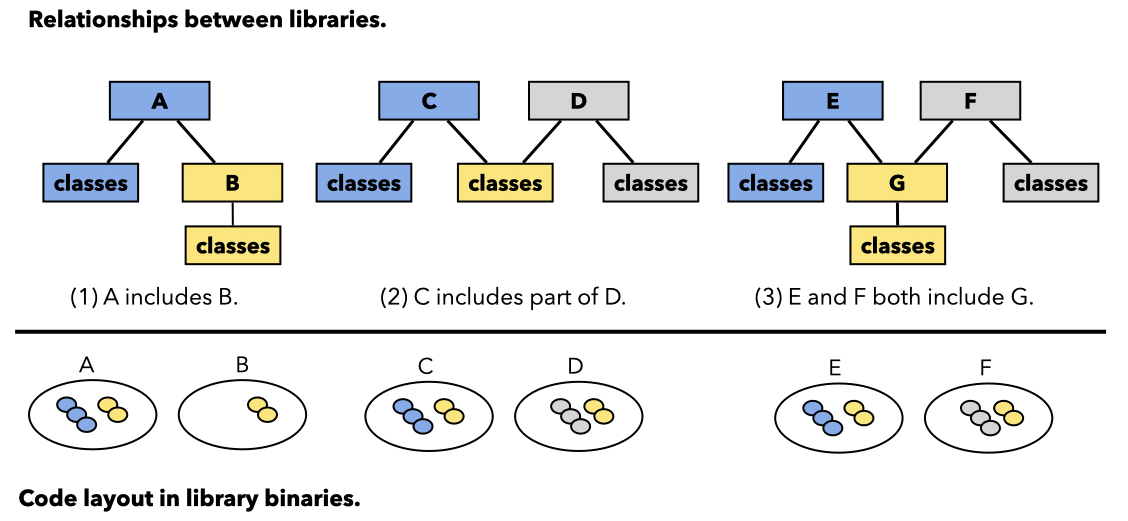}}
\caption{Three typical cases of code duplication (yellow region) between libraries.
}
\label{fig:relationship}
\vspace{-0.2in}
\end{figure}

By inspecting our collected libraries, we summarize the inter-library relationships into two categories, i.e., the \textit{library dependency} relationship, and the \textit{code sharing} relationship.

As described in Section~\ref{sec: integrating method}, libraries can integrate dependency libraries as apps do.
We denote this kind of relationship as \textbf{library dependency} and present three typical cases in Figure~\ref{fig:relationship}.
The differences locate in the inclusion degree (a complete inclusion like case 1, or a partial inclusion like case 2) and count of participants (two participants like case 1 and 2, or more participants like case 3).
Partial inclusion is possible when the library is customized or only sub-modules are integrated.

The other relationship \textbf{code sharing}
 refers to the existence of similar or identical code snippets between libraries. Significantly, these snippets do not exist as an independent library.
We can also use Figure~\ref{fig:relationship}(3) to illustrate a code sharing case. Here the $G$ node represents a shared code snippet instead of a library.
Some typical cases are summarized in the following:
\smallskip
\begin{itemize}[nosep,leftmargin=1em,labelwidth=*,align=left]
\item \textit{Sub-module packaging}\tab 
When a sub-module is packaged as an individual library, there is duplicated code between it and the master library. 
For example, the early Facebook-iOS-SDK was composed of sub-modules (CoreKit, LoginKit, and ShareKit) which were later encapsulated as libraries (FBSDKCoreKit, FBLoginKit, and FBShareKit). 
We collect multiple versions of them all and find a complex duplication situation among their binaries.

\item \textit{Library substitute}\tab 
To some extent, the deprecated library and its substitute can be regarded as one library's different versions. 
For example, GoogleMobileAds is deprecated in favor of Google-Mobile-Ads-SDK. 
There was no difference between the two when the handover just started. 
As the latter keeps updating, the gap gradually widens. 
But even so, there is still a lot of identical code between the two libraries.

\item \textit{Different releases}\tab 
The different releases are towards different developers. 
They have almost the same functionalities, and even the version updates are consistent. 
For example, the difference between AMapFoundation-NO-IDFA and AMapFoundation is whether the IDFA (identifier for advertising) is introduced.
\end{itemize}

Since Objective-C does not support namespaces, all classes in a binary (an app or a library) are organized and accessed in a flat space.
Therefore, we use scattered dots to represent classes inside a library to show the code layout, as presented in Figure~\ref{fig:relationship}.
For each library pair, when the app includes one of them, it seems similar to the other.

\subsubsection{Code duplication statistics}

\begin{figure}[t]
\centerline{\includegraphics[scale=0.35]{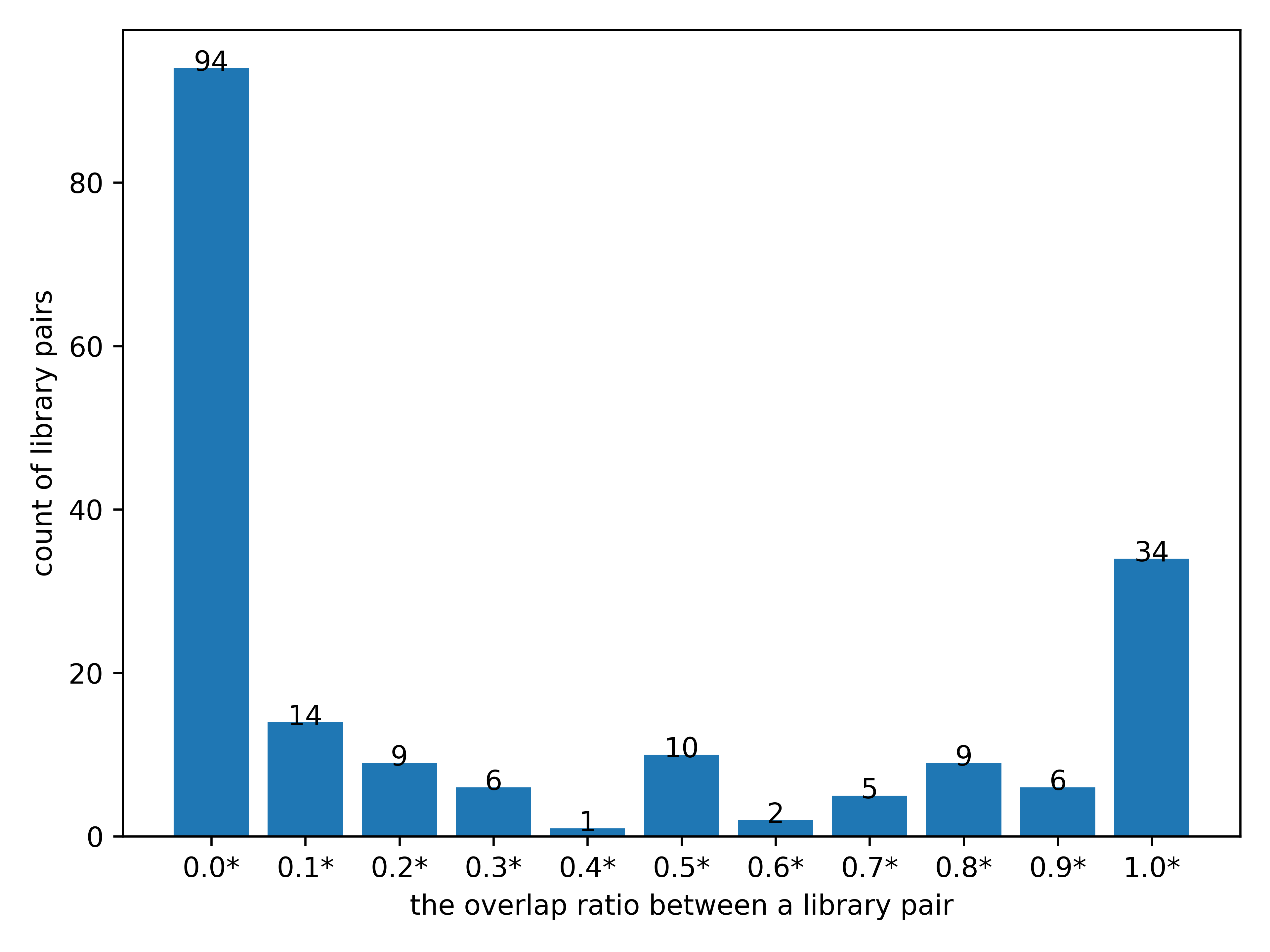}}
\caption{Library pairs with overlap regions.}
\label{fig:library overlap}
\end{figure}

To measure the code duplication in our collected library database, we calculate the overlaps between each library pair as follows:
\begin{equation}
overlap(A, B) = \max_{\forall a \in A, \forall b \in B} \frac{\#\ classes\ defined\ in\ a\ and\ b}{\#\  classes\ defined\ in\ a}, \label{0}
\end{equation}
$a/b$ is any collected version of library $A/B$. 
We regard the classes of the same name in $a$ and $b$ as their overlap region. 
If there is an overlap between $A$ and $B$, we keep both $overlap(A, B)$ and $overlap(B, A)$ since they have different meanings. 
For example, $overlap(YYKit, YYText)$ equals $0.34$ means that if an app includes the library $YYText$, we can also consider that the app includes $34\%$ of $YYKit$; on the other hand, the $overlap(YYText, YYKit)$ equals $0.87$ means that when the app includes $YYKit$, we can consider the app includes 87\% of $YYText$.

We present the distribution of overlap ratio among the collected $319$ libraries in Figure~\ref{fig:library overlap}.
The result shows that 84 libraries have code duplication with others.
A caution here is that a library pair whose overlap ratio equals $1.0$ does not mean they are identical since the granularity of overlap is very rough. 
Nevertheless, it does mean that they are highly similar (the measurement of library uniqueness is presented in Section~\ref{sec:evaluation}).  
Besides, the causes of overlaps we counted (i.e., overlaps in our database) are mainly copy-paste integration and code sharing. 
According to our collection strategy, when a target library integrates dependency libraries through management tools, we do not take these dependencies as part of it.

As such, when conducting a similarity comparison between the target app and libraries, an app code snippet can match with snippets of different libraries. 
We have to identify its provenance correctly to achieve accurate library detection.

\begin{figure}
\centerline{\includegraphics[scale=0.15]{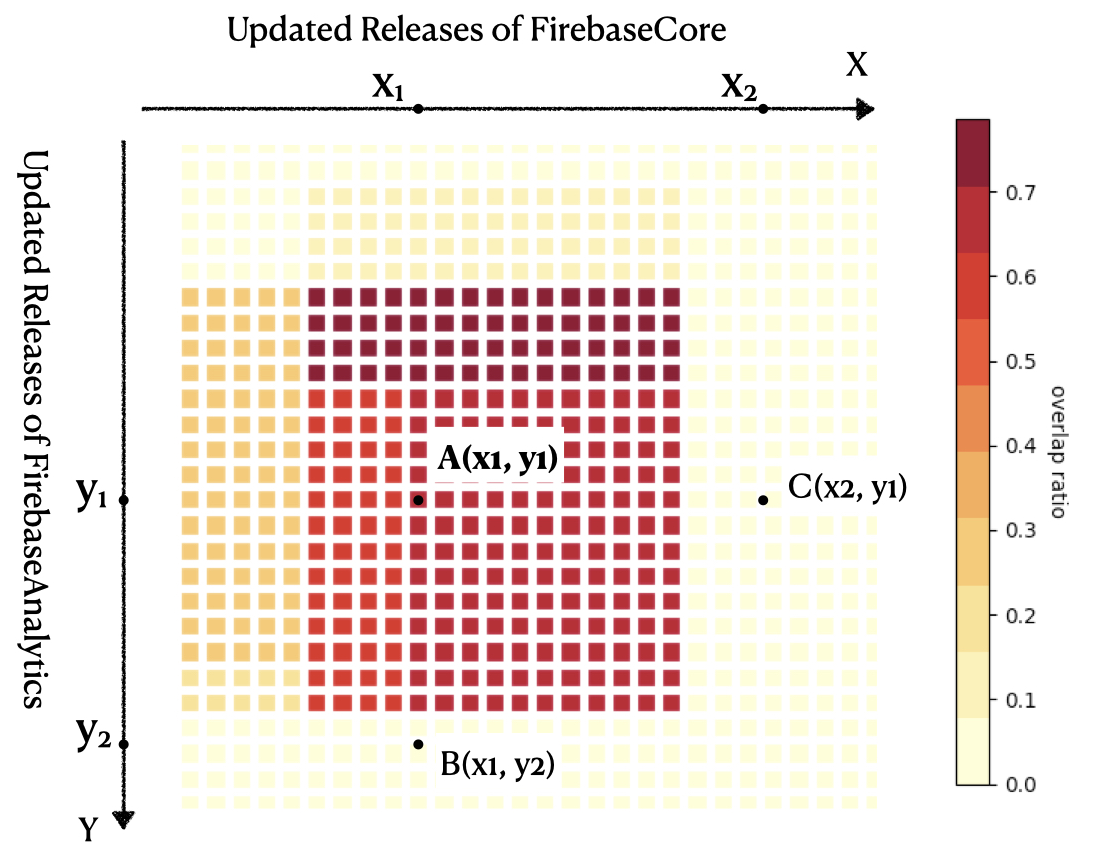}}
\caption{A partial map of the overlap ratio between FirebaseAnalytics and FirebaseCore in different version pairs. The color of each tile presents the extent of $overlap(FirebaseAnalytics_y, FirebaseCore_x)$, y/x is the specific releases of FirebaseAnalytics/FirebaseCore.
The overlap ratio is 0.6875 at point A, and 0 at point B and C. 
}
\label{fig:overlap map}
\end{figure}

\subsection{Our solution}

\smallskip \noindent\textbf{A motivating example}\tab
Figure~\ref{fig:overlap map} shows the overlap ratio between
two libraries, i.e., FirebaseAnalytics ($FA$) and FirebaseCore ($FC$) in different version pairs.
For a fixed version of $FA$, the overlap between it and different versions of $FC$ varied, since the later changes during updates.
For instance, at point $A$, $FA_{y1}$ has an overlap with the $FC_{x1}$. 
While at point $C$, the same $y1$ has no overlap with the updated $x2$ version of $FC$, which means $x2$ dropped the code that causes overlap at point $A$. 
This explanation also applies to point $A$ and point $B$. 
The duplicated (overlapped) code is removed from $FA$ after upgrading to a new version.
Since Objective-C does not have built-in support for namespaces, an app can only integrate a $FA$ as well as a $FC$ smoothly when this version pair has no overlaps.

Therefore, our solution to solve the challenge is based on two implications.
First, even though two libraries can share classes, the shared classes
cannot be integrated into an app at the same time without causing a class name conflict.
Second, code duplication between multiple versions of two libraries can vary. 

We take the library pair in Figure~\ref{fig:relationship}(2) to illustrate  our idea. 
For a given app, we first divide its classes into three groups, the $CommonClasses$, $CClasses$, and $DClasses$. 
For an app class $cls$, residing in $CClasses$ ($DClasses$) means it appears and only appears in some versions of library $C$ ($D$) 
while in $CommonClasses$ means it appears in some versions of library $C$ as well as some versions of $D$.
We assume that none of these three sets are empty.
We then use the union\{$CClasses$, $CommonClasses$\}
and union\{$DClasses$,$CommonClasses$\} to match with different versions of $C$ and $D$ to find a best match.
Note that, since each class in $CommonClasses$ can only belong to library $C$ or $D$ (implication one), we propose a strategy to decide which library (i.e., the $CClasses$ or $DClasses$) has priority to conduct the matching process.
If there is a match, then the library (and its version) is found.
Since code duplication between different library versions changes (implication two), there can be one version pair of $C$ and $D$ that do not share classes.
A more general algorithm is detailed in Section~\ref{sec:library instance recovery}.

\section{System Design and Implementation}
\label{sec:approach}

\begin{figure*}
    \centerline{\includegraphics[scale=0.25]{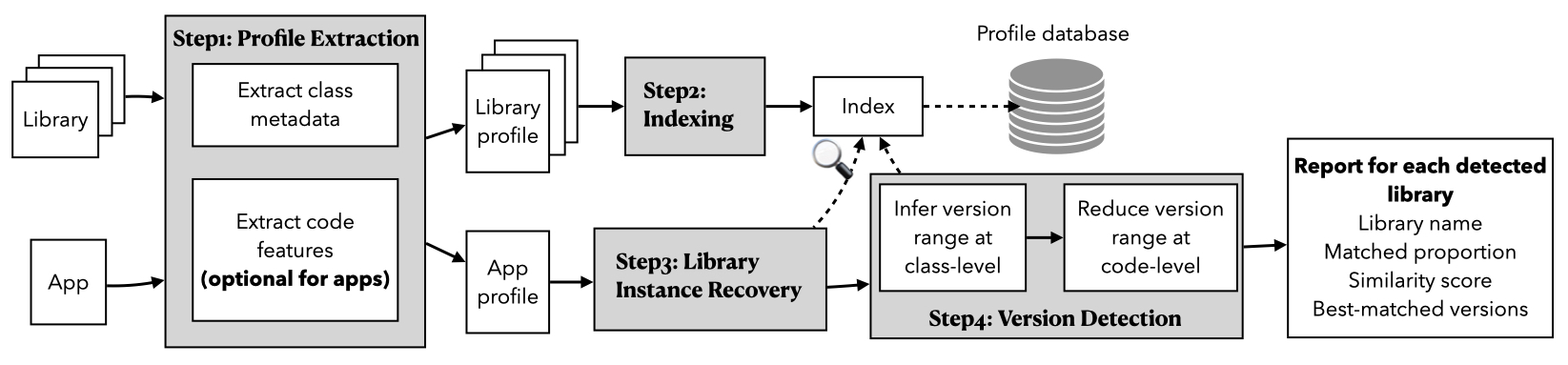}}
    \caption{Overall architecture of \tool{}.}
    \label{fig:approach}
\end{figure*}

\subsection{Overview}

Given an iOS app, our system aims at detecting the existence and version information of libraries included in it.
Accordingly, the system works in the following four steps, as shown in Figure~\ref{fig:approach}.


First, we build a profile for each library version we collected as well as the target app.
Second, with all the profiles of libraries, we build an index for them to facilitate searching.
Third, we recover library instances from the app based on its profile.
We handle the code duplication challenge in this step (Section~\ref{sec:library instance recovery}).
At last, we find out the best-matched version (or version set) for each recovered library instance.
It is worth noting that we devised a two-level profile structure, i.e., 
the \textit{class-level profile} and the \textit{code-level profile}.
The class-level profile is easy to build but less informative, 
while the code-level profile captures more variation between versions but takes time to construct. 
We recover library instances and infer their coarse version ranges with class-level profiles, and shrink the range using code-level profiles.
To make a trade-off between detection efficiency and version precision, we leave the code-level profile construction and fine-grained version detection for the target app as an optional task.
In the remaining part of this section, we will elaborate on each step in order.

\vspace{-0.1in}
\subsection{Profile Extraction}
\label{sec:profile extraction}

\begin{figure}[t]
\centerline{\includegraphics[scale=0.17]{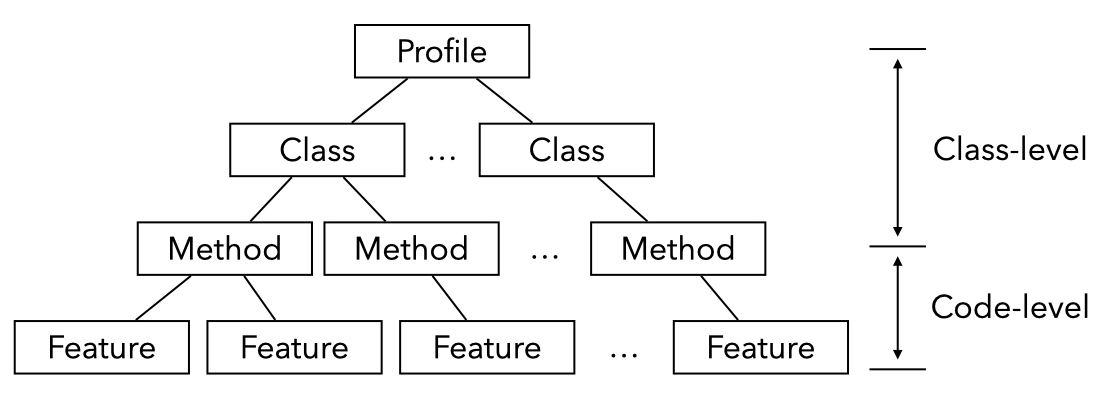}}
\caption{The universal profile structure for iOS apps and libraries. }
\label{fig:profile}
\vspace{-0.2in}
\end{figure}

\subsubsection{Profile structure}
We design the profile structure against the following observations.
\begin{itemize}[nosep,leftmargin=1em,labelwidth=*,align=left]
     \item \textbf{Obfuscation is not encouraged in iOS app.}\tab Apple does not encourage symbol obfuscation for iOS apps, which it believes will obstacle the review process~\cite{AppStore54:online}. 
     Generally, an app that contains obfuscated code will be declined by the reviewer, and the developer will be asked to explain the reason for the obfuscation. 
     Therefore, app developers rarely take the initiative to obfuscate third-party libraries. 
    \item \textbf{Class-level analysis of an iOS app is efficient.}\tab Due to the dynamic features of Objective-C, an iOS app written in this language needs to retain the metadata of all classes to support the Objective-C runtime.
    This metadata contains plenty of information and can be extracted easily by tools such as \code{class-dump}~\cite{nygardcl48:online}. 
    \item \textbf{Code-level analysis of an iOS app is time-consuming.}\tab Analyzing an iOS app is time-consuming. It usually takes IDA~\cite{IDAProH69:online} tens of minutes or even hours to only finish the initial auto analysis of an iOS app binary.  
\end{itemize}

\smallskip
Take both efficiency and accuracy into consideration, we design the profile with a two-level structure, as shown in
Figure~\ref{fig:profile}. 
Specifically, we construct the \textit{class-level profile with class metadata and the code-level profile with semantic features extracted from the assembly  after IDA's initial analysis}. 

\subsubsection{Class-level profile}
We define a virtual root node to help organize all classes included in a binary, and each class node has its defined methods as child nodes.
Due to the lack of namespaces, Objective-C classes must have unique names across an entire binary to avoid
the name conflict, and so do methods~\cite{Conventi40:online}. 
Therefore, each node in a profile has a unique label. 
During the extraction process, we tackle the following issues.
\smallskip
\begin{itemize}[nosep,leftmargin=1em,labelwidth=*,align=left]
    \item \textbf{Dump class metadata inside static libraries.}\tab 
    Since the class-dump tool cannot directly act on static libraries, we obtain a library's class metadata with the aid of an app containing it. 
    (Specifically, the Container app we obtained in the collection process. Check ~\ref{subsubsec:libdb} for more details.)
    We first dump the app's class metadata and then filter it with exported symbols of the target library if necessary.
    We use the nm~\cite{llvmnmli85:online} tool to dump the symbol table.
    \item \textbf{Deal with the class category.}\tab In Objective-C, the category is a structure to add methods to existing classes~\cite{Customiz69:online}. Different libraries and library versions can have separate extensions to an existing class, like \code{NSData(GMSCrypto)} and \code{NSData(FIRBase64)} are two categories of the class \code{NSData}.
    When building profiles, we deal with categories in the same way as classes, i.e., we will set \code{NSData(GMSCrypto)} and \code{NSData(FIRBase64)} as two distinct class nodes in a profile if they occur in one binary. 
\end{itemize}

\subsubsection{Code-level profile}


In this paper, we use the callee information in a method body as its code-level features.
As a dynamic language, Objective-C resolves method invocations (the messages) at runtime.
Specifically, each message is first delivered to the the dispatch function \code{\_objc\_msgSend} ~\cite{objcmsgS55:online,ObjectsC57:online}, then the dispatch function resolves the message and pass it to the real handler.
As such, a message is consists of three parts: the $receiver$'s pointer, the $selector$ for selecting the handler out from the $receiver$'s methods, and argument list for the handler.
Obviously, all those parts can be constructed dynamically at runtime, which complicates the call analysis.
Therefore, we directly pick the explicit data that helps decide the callee at runtime as code-level features.

We retrieve the features using the state-of-the-art tool IDA~\cite{IDAProH69:online}. 
After the IDA's initial automated analysis, we build a data reference graph through the interface provided by IDAPython to resolve the constant data usages in each method. 
Based on our observation, there are mainly four kinds of constant data that can help decide the identity of a callee—the class references (used to reference a class), the selector references (reference a selector), the constant strings (always used as parameters), and external symbols (mainly the data in \_\_stubs and \_\_got segments, including function symbols and data symbols). We record all these data usages in a method as its features list.




\subsection{Indexing}
To quickly retrieve classes, we unfold all class nodes contained in library profiles and index them by name. 
Given a name string, we can locate library classes of this name and access each class node's information defined in its profile, such as which library version it belongs to and what methods it defines. 
Such information are indexed and saved into database for efficient querying.

\subsection{Library Instance Recovery}
\label{sec:library instance recovery}
\begin{figure}
\centerline{\includegraphics[scale=0.15]{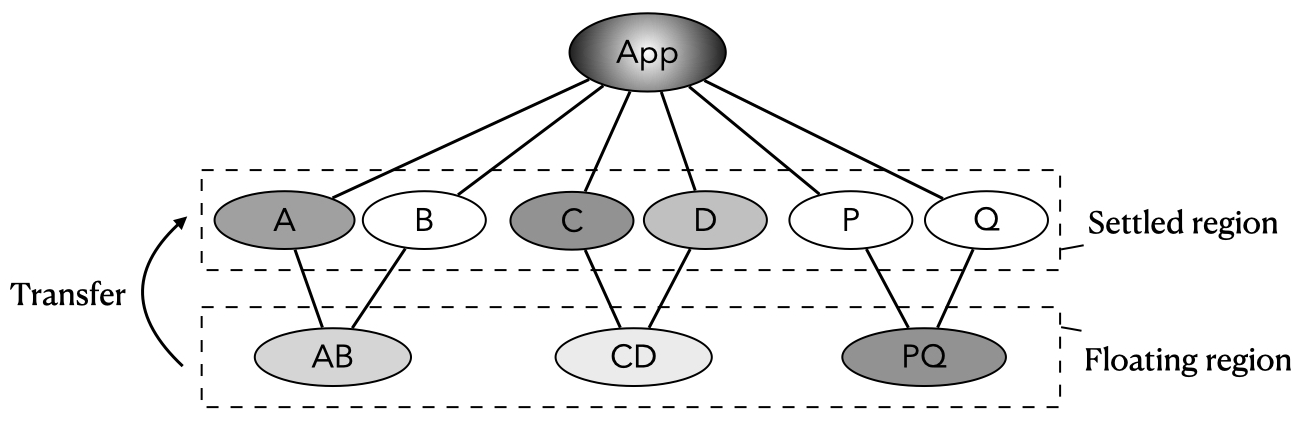}}
\caption{Arrange the app's classes by their possible third-party provenance. The gray-scale of a node represents the number of classes it contains.}
\label{fig:affiliation_graph}
\vspace{-0.2in}
\end{figure}

According to Section~\ref{sec:motivating}, an app profile can match with several library profiles at the same time due to the code duplication between libraries. 
Given the situation, we perform the match process at the class level instead of the library level. 
Even so, an app class can match multiple classes from distinct libraries or different versions. 
Therefore, to facilitate processing, we arrange app classes into a graph as shown in Figure~\ref{fig:affiliation_graph}. 
We put an app class to the floating region if it matches distinct classes of different libraries, otherwise, to the settled region.
For instance, a class locates in node $AB$ means it may be either library $A$ or library $B$ who introduces it into the app.
Suppose each node in the settled region is a library candidate, the floating classes can influence the judgments of candidates in two aspects.
\begin{itemize}[nosep,leftmargin=1em,labelwidth=*,align=left]
    \item The validness. If we award all classes of $AB$ to candidate $A$, then the candidate $B$ is spurious.
    \item The version. For a class in $CD$ that is introduced by $D$, awarding it to the wrong provenance $C$ may mislead the version judgment of candidate $C$. 
\end{itemize}

As a conclusion, identifying the correct provenance for floating classes is the critical point to recover accurate library instances. 
To achieve the goal, we recover library instances in two steps. 
We first arrange app classes by their possible provenance and construct an initial graph like Figure~\ref{fig:affiliation_graph}. 
Then we transfer classes in floating region to its best-matched candidate in the settled region.

\subsubsection{Identifying library candidates}
For each class $ac$ in the app, we compare it with all library classes with the same name to decide whether it belongs to some third-party libraries or not. 

\begin{itemize}[nosep,leftmargin=1em,labelwidth=*,align=left]
\item \textbf{STEP 1:} Locate library classes with the same name. 
    To avoid naming conflicts with other libraries or app-authored code, library developers try their best to assign special prefixes to their classes. 
    Therefore, we use the class name to locate library classes that may be similar to $ac$ and denote them as set $S_t$.
\item \textbf{STEP 2:} Find classes similar to $ac$. 
    For each library class $lc$ in $S_t$, we calculate the similarity score between $ac$ and $lc$ as follows:
    \begin{equation}
    sim(ac, lc) = \frac{\#\ methods\ defined\ in\ ac\ and\ lc}{\#\ methods\ defined\ in\ ac\ or\ lc}. \label{1}
    \end{equation}
    We regard all $lc$s whose $sim(ac, lc)$ is greater than 0 as the counterparts of $ac$ and denote them as $cp(ac)$. 
\item \textbf{STEP 3:} Award $ac$ to the appropriate node in the class layout graph. 
    Suppose $ac$ and its counterparts are as follows (the digit is the similarity score):
    $$ac: lc_1(0.8), lc_2(0.8), lc_3(0.7), ...$$
    If all these counterparts are from library $A$ (but different versions), award it to node $A$; if part of them are from $A$ and others are from $B$, then award it to node $AB$.
\end{itemize}

\subsubsection{Filtering library candidates}

In this step, we move each floating class to the library candidate that it most likely belongs to. 
We present the detailed transfer strategy in Algorithm~\ref{algo:filter}. 
Before that, we first introduce the assumptions our strategy based on, and then detail the indicators that help us make decisions later.

\smallskip \noindent\textbf{Assumptions}\tab

The first assumption we make is that the library candidate with more classes has priority to match against floating classes.
For example, for a candidate pair like $A$ and $B$ in Figure~\ref{fig:affiliation_graph}, the app could include $A$ and $B$ both, or include $A$ only. This all depends on the provenance of floating classes in $AB$.
Since $A$ contains more classes than $B$, which means it contains more information to help decide if a floating class in $AB$ belongs to it, we start from $A$ to match it against each floating class in $AB$.

The second assumption is that when a floating class can co-exist with a library candidate, we consider this floating class is compatible with the candidate and should be part of it.
Specifically, for a class $cls_f$ in $AB$, to decide whether it belongs to $A$ or not, we calculate the version range for current $A$ based on classes in it and the version range for $cls_f$.
If the two ranges have an intersection, we consider they can coexist in one binary (version) and move $cls_f$ to candidate $A$.

However, these two assumptions can not handle all cases.
For instance, for candidate pair $P$ and $Q$ who does not carry any information to match against classes in $PQ$, or a floating class in $AB$ failed to match with $A$ while $B$ carries no information, or a floating class in $CD$ failed to match with both $C$ and $D$, we take a trial and error.
We give each candidate a try to own such floating classes and calculate a $Score$ based on the current state, and then award the floating classes to the candidate with a higher score. 
The $Score$ is version sensitive, thus awarding fake classes to a candidate can make the candidate abnormal and results in a low score. We denote this as assumption 3.

\smallskip \noindent\textbf{Indicators}\tab
For a library candidate $C$, $L$ is its corresponding library, $V$ is a set of collected versions of $L$ and $V_x$ is one of them, we define the following indicators.
\begin{itemize}[nosep,leftmargin=1em,labelwidth=*,align=left]
    \item $M(C, V_x) = \{ac \mid ac \in C, cp(ac) \cap V_x \ne \emptyset \}$, the matched classes between $C$ and $V_x$.
    Classes in $C$ are detected to belong to $L$, but they do not necessarily exist in $V_x$. 
    In the worst case, there is no $V_x$ contains all these classes—if the integrated $L$ is a missed version, or $C$ includes some wrong classes by mistake.
    \item $Sim_s(C, V_x) = \sum_{ac \in M(C, V_x)}sim(ac, lc_v)$, the aggregate similarity score between $C$ and $V_x$. For each $ac$ in $M(C, V_x)$, $lc_v$ is its counterpart in $V_x$, and $sim(ac, lc_v)$ is the similarity score between them, which we have calculated before when finding library classes similar to $ac$.
    \item $Sim_a(C, V_x) = {Sim_s(C, V_x)}/{\#\ M(C, V_x)}$, the average similarity of the matched classes between $C$ and $V_x$.
    \item $Prop(C, V_x) = {\#\ M(C, V_x)}/{\#\ V_x}$, the matched proportion in the original library version.
\end{itemize}
Based on these objective indicators, we introduce three subjective attributes defined by ourselves. 
\begin{itemize}[nosep,leftmargin=1em,labelwidth=*,align=left]
    \item $Comp(C, V_x) = {\#\ m(C, V_x)}/{\#\ C}$, the compatibility between $C$ and $V_x$. As we explained in the definition of $M(C, V_x)$, this value can be smaller than 1. In this case, the more classes there are in $M(C, V_x)$, the more compatible the $C$ is in $V_x$.
    \item $V_p = \argmax_{V_x \in V}{Sim_s(C, V_x)}$, the best matched version set. Since there may be class additions and deletions between versions, we use the aggregate similarity to decide $V_p$ instead of the average. Besides, we do not consider $Prop$ when deciding $V_p$ due to the possibility of partial inclusions.
    \item $Score(C) = \max_{V_x \in V_p}Sim_a(C, V_X)*Prop(C, V_x)*Comp(C, V_x)$, the final score we assign to candidate $C$.
    The similarity score and matched proportion are the two general factors to measure whether an app and a library match. 
    Here we use an additional $Comp$ to magnify the impact of placing a class into a candidate by mistake. 
\end{itemize}

\algsetup{linenosize=\small}
\SetAlFnt{\small}
\begin{algorithm}[t]
\caption{Filter library candidates.}
\label{algo:filter}

\DontPrintSemicolon
\KwIn{G: the region graph.}
\KwOut{R: valid library instances}
$R \gets \emptyset$\;
$N \gets $ \textit{Non-empty nodes in settled region}\;
\For{$lib \ in \ sorted(N, key=size, reverse=True)$}{
  \For{$fnode \ in \ lib's\ successors$}{
    \For{$ac \ in \ fnode$}{
      \If{$ac\ is\ compatible\ with\ lib$}{
        \textit{move ac from fnode to lib}\;
      }
    }
  }
}
\;
$N \gets $ \textit{All nodes in settled region}\;
\For{$lib \ in \ N$}{
  $extended\_lib \gets lib\ +\ lib's\ successors$\;
  $lib.score \gets extended\_lib.score$\;
}
\While{N is not empty}{
  $lib \gets sorted(N, key=score, reverse=True)[0]$\;
  \textit{N.remove(lib)}\;
  \If{$lib.score > 0$}{
    $R.add(lib)$\;
    \For{$fnode \ in \ lib's \ successors$}{
      \textit{move all classes contained in fnode to lib}\;
      $del \ \langle lib,\ fnode \rangle \ in \ G$\;
      \For{$rlib\ in \ fnode's \ predecessors$}{
        $update\ rlib's\ score$\;
      }
      }
    }
  }
  \Return R\;
\end{algorithm}
\smallskip \noindent\textbf{The algorithm}\tab
According to assumption 1, we start with non-empty candidates and rank them by size (i.e., the count of classes in it).
Then, for a candidate, we check each floating class in its descendants to see whether this class is compatible with it based on assumption 2.
Specifically, we consider a candidate and a floating class as compatible if their $V_p$s have an intersection.
When calculating $V_p$ for a single class, we regard it as a candidate that has only one class in it.
As a result, we transfer the floating class to the candidate if they match, or we leave it in the floating region waiting for its next turn.
Since awarding a class to a candidate may change the latter's $V_p$, we keep updating $V_p$ for each compatibility test.
We present this part in lines 3-7 of Algorithm~\ref{algo:filter}.

After the first round transfer, some floating classes may have settled down, while some may remain floating. 
As such, we make a trial and error—let each candidate try owning all possible classes and check how do they fit. 
As presented in lines 9-12, we assume every candidate owns all classes in its descendants and calculate the $Score$ for the extended one. 
According to assumption 3, we rank candidates by their extended $Score$ and award floating classes to the candidate with a higher $Score$. 
We present the detailed logic in lines 13-22. 
It's worth noting that, when we confirm the provenance of a floating class, we have to remove it from some other candidates that own it for the moment and update their scores, as presented in lines 21-22. 
In the end, we regard each returned library candidate as valid library instance and recalculate indicators for it, which will be output as part of the report.

\vspace{-0.1in}
\subsection{Version Detection}
Since we have calculated $V_p$ for each library instance in the previous step, we take it as the class-level version detection result. 
If $V_p$ is not as precise as expected, we regard it as candidates and conduct a further detection based on code-level profiles. 
Considering that differences between neighboring versions are usually small, and the candidate versions are often successive, the code involved in differences between candidates is limited. Focusing on these code snippets, we can find the best-matched versions efficiently.

Since we extract code features on method granularity, we first find methods that are not consistent among candidates. 
To be precise, the methods exist in all candidates but inconsistent among them, and methods that exist in part of candidate versions. 
We denote these methods as set $M$. 
Meanwhile, we denote the intersection set between the library instance and $M$ as $N (N\subseteq M)$. 
We define function $f(m, B)$ to obtain the code features of method $m$ in binary $B$, where $B$ can be a library instance or any candidate version. 
When there is no $m$ available in $B$, an empty feature list is returned. 
Therefore, the similarity score between a library instance $L$ and a candidate version $C_i$ can be expressed as:
$$sim(L, C_i) = \sum_{m \in N}sim(f(m, L), f(m, C_i)).$$

We use a variant of Manhattan distance to measure the similarity of two feature lists. This method is used for clone detection in Wukong~\cite{wang2015wukong}. For feature lists A and B, with n kinds of features in total, their similarity score is represented as:
\begin{equation}
sim(A, B) = 1-distance(A, B) = 1-\frac{\sum_{i=0}^{n}\lvert A_i-B_i\rvert}{\sum_{i=0}^n(A_i+B_i)}.\label{5}
\end{equation}
Finally, the candidates with the highest similarity score are given as the final result.

\section{Evaluation}
\label{sec:evaluation}

Our evaluation is driven by the following three research questions: 

\begin{itemize}
    \item[RQ1] Can the profiles we extracted distinguish different libraries and library versions?
    \item[RQ2] Can \tool recover library instances from apps and identify their versions correctly? 
    Does it handle the code duplication between libraries well?  
    \item[RQ3] Can \tool facilitate the detection of vulnerable library versions?
\end{itemize}
 
To answer RQ1, we measure the profile uniqueness across the collected libraries.
To answer RQ2, in the absence of established benchmarks for iOS third-party library detection in the research community, 
we propose to collect open-source apps with known library usage as a benchmark. 
To answer RQ3, we focus on three well-known vulnerable libraries and perform a large-scale detection in the wild.
We intend to make our data publicly, including the \tool project and the dataset.

\subsection{Dataset}

\subsubsection{Library collection}
\label{subsubsec:libdb}
To automate the collection process, we download library repositories via CocoaPods~\cite{CocoaPods} and then compile them using Xcode~\cite{Xcode12A12:online}. 
To make our collected libraries representative, we use Specs Repo~\cite{CocoaPod40:online} to obtain the popularity of libraries.
Specs Repo is a repository that maintains records of each library hosted on CocoaPods, including the name, version, source, license, summary, dependencies, etc.
Therefore, we can measure a library's popularity by the frequency of being cited as a dependency.
After confirming the target library list, we try to collect as many versions as possible for each target. 
Significantly, when a target library has dependencies, we add the dependency libraries to the target list also.
Each library version is collected as follows.

\begin{itemize}[nosep,leftmargin=1em,labelwidth=*,align=left]
\item[1)] Create an empty Xcode project (\textit{the Container app}) for integration.
\item[2)] Specify the target library's name and version in a Podfile. CocoaPods will first analyze its dependency graph, and then integrate all of them into the Container app. Meanwhile, a tracking file (\textit{the Podfile.lock}) will be generated to record \textit{the dependency graph between libraries as well as each installed library's name and version.}
\item[3)] Use the Xcode toolchain to build the Container project.
\item[4)] Dump library data and clean the Container project. For each library version, we collect three items: \textit{binaries of the target library (the released original binaries or the compiled static library), the Container app, and the Podfile.lock file}. We will use them in later tasks. 
\end{itemize}

In total, we collect 319 unique libraries, with 5,768 versions.


\subsubsection{Benchmark}
As aforementioned, when an app integrates third-party libraries through CocoaPods, the CocoaPods will
generate a Podfile.lock file to record all details of this integration, including each
library's name and version. Therefore, we collect apps that carry Podfile.lock files to build the benchmark for testing.

The whole benchmark consists of two app sets. 
The first is container apps obtained during library collection. As a side product of
library collection, each container app includes a target library and its dependency libraries (if any).
The second is real-world apps selected from the open-source-ios-apps project~\cite{opensour88:online}, 
since we can only obtain Podfile.lock files from open-source repositories. 
This set contains 39 real-world iOS apps with $501$ valid library integrations through CocoaPods.

\subsection{RQ1: Profile Uniqueness}
\label{sec:profile evaluation}
\begin{figure}
\centerline{\includegraphics[scale=0.3]{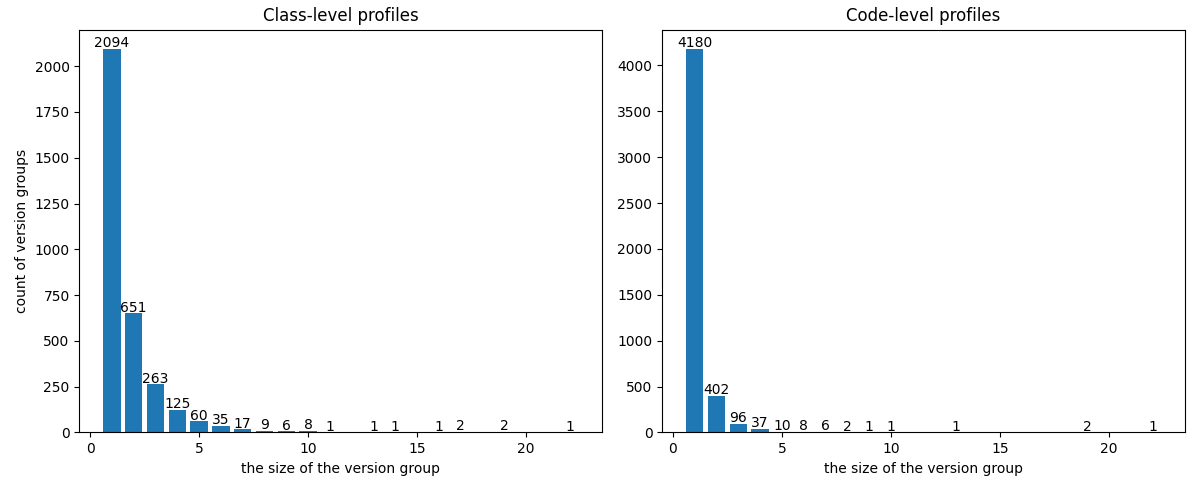}}
\caption{
    The uniqueness of library profiles. Versions in a group have the same profiles.
}
\label{fig:profile_uniqueness}
\vspace{-0.2in}
\end{figure}

Among the 5,768 versions of 319 libraries, we find 100 library versions whose profile is empty. 
Most of them carry no code snippets and provide functions by integrating dependency libraries; 
the others contain valid code regions, but our extraction process failed due to their unexpected code organizations.
As a result, we have 5,668 versions of 318 libraries to support the uniqueness evaluation. 

To facilitate the evaluation, we calculate two signatures for each profile. One is calculated based on its class-level content, and the other on the code-level. 
We group all those 5,668 profiles by their signatures so that each group contains profiles with the same class-level or code-level signature.
The grouping results are presented in Figure~\ref{fig:profile_uniqueness}.
When an app integrates any library version inside a group, \tool will identify the whole group as candidate versions. 
According to Figure~\ref{fig:profile_uniqueness}, 87.9\% (4985/5668) of our class-level profiles can limit the candidate versions to 5 ($\le 5$), 
and 96.5\% (5470/5668) of code-level profiles can achieve the same result.


In most cases, profiles distinguish between libraries well, except for three cases. 
We inspect their binaries and list the reasons below.

\begin{itemize}[nosep,leftmargin=1em,labelwidth=*,align=left]
\item \textit{WechatOpenSDK v1.8.6 and TMWechatOpenSDK v0.0.1, v0.0.2.} We find their binaries are consistent by diffing them. According to the latter's self-description on CocoaPods, it is an encapsulation of the former library; and according to the profile signatures, we can infer that the two TMWechatOpenSDK versions encapsulated WechatOpenSDK v1.8.6 exactly. 
\item \textit{AMapLocation and AMapLocation-NO-IDFA.} Although we can infer that there must be a difference between them according to names, binaries of the same version of the two libraries are consistent. They are different because they rely on different libraries, the former integrates AMapFoundation while the latter includes AMapFoundation-NO-IDFA. 
\item \textit{Google-Mobile-Ads-SDK and GoogleMobileAds.} The latter is deprecated in favor of the former. The latter stop release since v7.8.0, but the former has been released for some time before v7.8.0. During the period of joint release, the libraries of their corresponding versions may be highly similar, at least the v7.7.1 of the two libraries are consistent at the binary-level.
\end{itemize}


Taking both the general update content and the profile structure into consideration, our profiles \textit{cannot} distinguish between versions when the updates all located in the following regions: 1) files without valid codes, such as README, comments, resources, test codes, etc.; 2) valid code region but beyond our profiles, such as variables, C/C++ functions, swift methods, NSConcrete*Blocks; 3) and changes related to control flow in Objective-C methods.  

\begin{framed}
\noindent \textbf{Answer to RQ1:} 
\textit{
Except for a few exceptions, the extracted profiles distinguish between different libraries well, even at the class-level. 
Regarding the discernibility between versions, 87.9\% of class-level profiles can limit the candidate versions to 5 ($\le 5$), and 96.5\% of code-level profiles can achieve the same result.
}
\end{framed}

\subsection{RQ2: Library Instance Recovery and Version Detection}

\begin{table}[t]
\caption{Results of library instance recovery.
}
\begin{center}
	 \resizebox{0.8\linewidth}{!}{%
\begin{tabular}{ccccc}
\hline
App Set & Precision & Recall &\# Library Uses & \# TP \\
\hline
Containers & 0.991 & 0.998 & 9994 & 9978 \\
Real-world apps & 0.977 & 0.998 & 501 & 500 \\
\hline
\end{tabular}
}
\label{tab1}
\end{center}
\end{table}

\begin{table}[t]
\caption{Results of library version detection. Suppose the detected version set is $S_d$ and the correct version is $v$, C, S, I, refers to Correct ($S_d=\{v\}$), Sound ({$\{v\} \subset S_d$}), and Incorrect ($v \notin S_d$).
}

 \resizebox{0.8\linewidth}{!}{%
\begin{tabular}{@{}llllllll@{}}
\toprule
\multirow{2}{*}{App Set} & \multirow{2}{*}{\# Total} & \multicolumn{3}{l}{After Class-level} & \multicolumn{3}{l}{After Code-level} \\ \cmidrule(l){3-8} 
                &      & C(\%) & S(\%) & I(\%) & C(\%) & S(\%) & I(\%) \\ \midrule
Containers      & 9247 & 38.8  & 60.6  & 0.5   & 73.6  & 25.2  & 1.3   \\ \midrule
Real-world apps & 461  & 43.4  & 54.4  & 2.2   & 70.5  & 23.4  & 6.1   \\ \bottomrule
\end{tabular}
}
\label{tab2}
\end{table}

To evaluate our approach, we \textbf{first} labeled $9,994$ library uses in 5,768 container apps and $501$ in 39 real-world apps.
Each library use has a record in its corresponding Podfile.lock file. 
Besides, we can find proofs of this integration, such as a partially matched class, to ensure the library use is genuine and detectable.
\textbf{Then}, we applied \tool to recover library instances from apps. 
We denote a recovered instance as a $TP$ (true positive) if we have labeled it, an $FP$ (false positive) if not. 
All library uses that \tool missed are denoted as $FN$s (false negatives). 
\textbf{Next}, we conducted a class-level version detection for each $TP$ whose correct version ($V_c$) we have collected. 
To facilitate evaluation, we mark a detection result (a version set) as $correct$ if it contains $V_c$ only, $sound$ if it contains more than $V_c$ and $incorrect$ if it does not include $V_c$.
\textbf{Last}, we conducted code-level version detection for each $TP$ whose class-level output is $sound$.
Notably, we accept the case a WechatOpenSDK v1.8.6 library use was detected as a TMWechatOpenSDK v0.0.1 since they have identical profiles (see Section ~\ref{sec:profile evaluation}).

We present the results of instance recovery and version detection in Table~\ref{tab1} and Table~\ref{tab2} respectively. 
According to Table~\ref{tab1}, \tool achieves a good precision and recall in library recovery for both the two datasets.
A precision of 99.1\% and a recall of 99.8\% indicate that \tool rarely moves classes from real candidates to spurious ones, and our transfer strategy is effective. 
While for real-world apps, the recall remains 99.8\%, but the precision drops to 97.7\%.  We inspected the failed cases and speculate that the majority is due to copy-paste integrations or unacquainted integrations. It is hard to distinguish whether the app includes a partial library that we have collected or an uncollected library that overlaps with some collected ones. In any case, this is the best result we can give based on the current database.

The accuracy of version detection is reasonable when considering the inherent uniqueness of library profiles. 
In addition to the reason that library profiles are not sufficient to distinguish between versions, the reason for those $sound$ results may also be partial inclusions, and the included part is consistent between the output versions.
Besides, there is a sharp increase of $incorrect$ results for real-world apps. 
By inspecting those $incorrect$ cases, we found that most of them have been customized before integration, which interferes with the version detection process.

Next, we discuss some reasons why \tool cannot work as expected. We inspected the $incorrect$ version detection results at code-level and found most of them are due to compilation or decompilation diversity. 
Thus, we mainly analyze the reasons for mistakes in class-level phases, including recovering library instances and detecting versions at class-level.

\begin{itemize}[nosep,leftmargin=1em,labelwidth=*,align=left]
\item \textit{The app integrates an uncollected library $A$, and we collected a library $B$ that overlaps with it.} If the app does not integrate $B$, there will be an FP (we identify out $B$); if the app integrates $B$ but not all versions of $B$ overlap with $A$, then there could be a $incorrect$ result in version detection for $B$ since we regard the $A$ code as part of the recovered $B$ instance. 
A similar case is we did not collect a library in its entirety. Some libraries may have sub-modules or extensions but require additionally specify them in the Podfile to integrate them. We may collect only the modules installed by default when collecting a library.

\item \textit{We do not generate a complete profile for some library versions.} According to Section~\ref{sec:profile extraction}, we filter class metadata by exported symbols of library binaries. If version $a$'s symbols are partially stripped while $b$ keeps them all, then we get a broken profile for $a$ and a complete for $b$. When an app integrates $a$, we identify it as $b$ since the recovered library instance is more similar to $b$.

\item \textit{Compilation diversity.} The diversity occurs in dealing with categories (see Section~\ref{sec:profile extraction}). The compiler may merge the contents of multiple categories of the same class into one category during linkage. However, moving the content in $Cat_a$ to $Cat_b$ will make $Cat_a$ seem to disappear in the binary, and $Cat_b$ to contain more methods. 
Since this difference not only occurs between apps and libraries but also between collected versions, it can affect deciding the provenance of an app class, as well as judging the version of a recovered library instance.
\end{itemize}

\begin{framed}
\noindent \textbf{Answer to RQ2:} 
\textit{
    From the evaluation results of library instance recovery and version detection, \tool handles the code duplication well. It achieves a recall exceeds 99\%, and a precision exceeds 97\% in library instance recovery. The version detection result is also reasonable according to the intrinsic uniqueness of library profiles.
}
\end{framed}

\subsection{RQ3: Vulnerable Library Detection}
To demonstrate potential usages, we apply \tool to detect vulnerable libraries in iOS apps. 
We select three Objective-C libraries as targets, AFNetworking, SSZipArchive, and GCDWebServer. These three libraries all have had defective versions and then patched them. 
Specifically, the v2.5.1 and v2.5.2 of AFNetworking are at risk of man-in-the-middle (MITM) attack~\cite{CVECVE2081:online, Arespons14:online}, and the last two libraries are suffering from a directory traversal vulnerability in some early versions~\cite{Releasev85:online}~\cite{CVECVE2060:online, Release32:online}. 
For each library, we have collected both the vulnerable and safe versions.

We tested $4,249$ real-world apps in total and obtained a result as shown in Table~\ref{tab:vulnerable library detection} (only class-level version detection was applied). 
A detected library was marked as $vulnerable$ if the detected version range contains vulnerable versions only, $safe$ if it contains safe versions only, and $risky$ if both vulnerable and safe versions are covered. 
Regarding the $risky$ results, there are mainly two reasons.
First, the recovered library instance is a partial library, and the recovered part is consistent among a broad range of versions—including the vulnerable ones and the safe ones. 
Second, the recovered library instance is a complete library, but the class-level profiles of vulnerable versions are consistent with some nearby safe ones.

Since we conducted class-level version detection only, it takes seconds to finish testing an app, which proves our tool is adaptive for extensive testing.
Besides, in addition to those three libraries, we can combine \tool with some dependency-check tools to check whether the target app contains publicly disclosed vulnerabilities.
For example, the CocoaPods Analyzer under the OWASP Dependency-Check project takes a Podfile.lock file as input and check the security of library dependencies contained in it~\cite{OWASPDep57:online, dependen60:online}. 
With \tool, we can reconstruct the Podfile.lock file for an app based on the detection result and feed it to the analyzer.
Although this analyzer is considered experimental for the moment, there is an application prospect.


\begin{table}[t]
\caption{The results of vulnerable library detection in 4,249 apps.}
\begin{center}
\begin{threeparttable}
	 \resizebox{0.8\linewidth}{!}{%
\begin{tabular}{lcccc}
\hline
\multirow{2}{*}{Library} & \multirow{2}{*}{Vulnerable versions} & \multicolumn{3}{c}{Detected libraries in apps} \\ \cline{3-5} 
             &                  & vulnerable\tnote{1} & risky\tnote{2} & safe\tnote{3} \\ \hline
AFNetworking	& $2.5.1, 2.5.2$	& 20	& 8	& 1446 \\
SSZipArchive		& $\leq 2.1.3$		& 324	& 79	& 88   \\
GCDWebServer	& $\leq 3.5.1$		& 61	& 62	& 11    \\ \hline
\end{tabular}
}
\begin{tablenotes}
	\footnotesize
	\item[1] The output version range contains vulnerable versions only.
	\item[2] The output version range contains both vulnerable and safe versions.
	\item[3] The output version range contains safe versions only.
\end{tablenotes}
\end{threeparttable}
\end{center}
\label{tab:vulnerable library detection}
\end{table}

\begin{framed}
\noindent \textbf{Answer to RQ3:} 
\textit{
Our tool can quickly identify third-party libraries included in apps and give its version range. Combining the answers of RQ1 and RQ2, the auxiliary ability and practicality in security tasks are promising.}
\end{framed}

\section{Discussion}
\label{sec:discussion}

The first limitation of our method is that we use version information to deal with code duplication, which in turn places high demands on the library collection. In theory, the more libraries and library versions we collect, the more accurate our method is. That's because when judging the provenance of code through compatibility, a large number of compatible cases are required as prior knowledge. This also provides us with another idea. We can judge the compatibility between a code fragment and a library through machine learning once we have an exhaustive database.

The second limitation is the profile content. As presented in the evaluation part 
there is still room for improvement in the soundness of profiles. 
Our future interest mainly lies in two aspects. The first is the diversity of iOS third-party libraries.
Except for Objective-C, there are also Swift and C/C++ developed libraries. 
The second is supplementing Block~\cite{Introduc39:online} into code features, which is a critical structure to achieve GCD in iOS and takes a considerable portion in the \code{\_\_text} section.
The version updates may occur in the Blocks. The challenge is a Block exists as an anonymous function in binaries, it has to be associated with non-anonymous functions to identify its identity. 

Third, we want to discuss the transfer strategy, although in most cases it works well in the light of evaluation.
First of all, according to assumption 1, library candidates with more classes have priority to match against floating classes.
However, when the classes in it are weak in version distinguishment, a wrong floating class can pass the compatibility test easily.
The second is the granularity of processing floating classes. 
In the first round, the processing unit is class; 
while in the second round, the unit is the class group. 
A single class may contain too little information thus results in a broad $V_p$, while a class group may be too large since the provenance of classes in it can be different theoretically. 
It will be better if we can first group floating classes inside a node into several small clusters by their internal compatibility.

\section{Related Work}

\subsection{Third-Party Library Identification}
Although detecting third-party libraries has been a common task in security research on iOS apps~\cite{egele2011pios,han2014android,chen2016following,wang2018software,tangios}, there are rare systematic studies targeting detecting iOS third-party libraries. As the only one, CRiOS~\cite{orikogbo2016crios} excavates third-party classes from a massive app set and group them into clusters (libraries). CRiOS first group classes by class name prefix, which is a weak alternative to namespaces in iOS development, then merge these clusters by class cohesion. Due to the intra-library relationships, the second step can make the obtained library cluster contains classes from more than one library. 

As another major mobile operating system, third-party library detection techniques targeting the Android platform continue to emerge~\cite{grace2012unsafe,book2013longitudinal,narayanan2014addetect,crussell2014andarwin,chen2014achieving,liu2015efficient,wang2015wukong,ma2016libradar,backes2016reliable,li2017libd,glanz2017codematch,zhang2018detecting,wang2018orlis,zhang2019libid}. Not only for the market demand but also for the resilience to the growing obfuscation techniques. However, due to the inherent differences between platforms, these methods are not well applicable to iOS.

Another line is identifying third-party packages in binaries~\cite{hemel2011finding,duan2017identifying,tang2020libdx}, mainly targeting binaries developed by C/C++. These methods extract string literals from binaries to characterize a package and conduct detection based on a large-scale database. 
However, this will result in huge duplicated features due to the string constants repeated in multiple libraries. 
Meanwhile, since code duplication is a common issue in open source projects~\cite{lopes2017dejavu}, they face the same code duplication challenge.
To address this challenge, OSSPolice~\cite{duan2017identifying} utilizes the structurally rich tree-like layout of OSS sources to deal with internal code clones (i.e., nested third-party OSS clones). 
However, we do not assume a library is organized as expected, and the duplicated code is in a typical nested relationship with master libraries.
On the other hand, LibDX~\cite{tang2020libdx} categorizes the code clone into parent-child nodes and library variants but misses the case that libraries sharing code snippets but outside the two relationships. Besides, it does not collect multiple versions for a single library, thus faces a less complicated code duplication problem as us.
\vspace{-0.1in}
\subsection{Security of iOS third-party libraries}
iOS third-party libraries play a critical role in many iOS security studies. 
During the research of privacy leakage in iOS apps, Egele et al.~\cite{egele2011pios} found 55\% of tested apps include either advertisement or tracking libraries. 
Moreover, all these libraries triggered PiOS' privacy leak detection since they transmit the unique device ID to third parties. 
By the time of Chen et al.~\cite{chen2016following}'s research on PhaLibs, there was still no available techniques to recover libraries in iOS apps, and also no available public anti-virus (AV) systems to analyze iOS third-party libraries. 
They utilize the associations between iOS libraries and their counterparts of the Android platform to find and analyze iOS PhaLibs. 
As a result, they found that most Android-side harmful behaviors were preserved on their iOS counterparts. 
In a recent work of vetting and analyzing network services of iOS apps, Tang et al.~\cite{tangios} identified the third-party network service libraries from apps based on a call stack similarity analysis and found that the use of certain third-party libraries listening for remote connections is a common source of vulnerable network services.

\section{Conclusion}
In this paper, we work towards the detection of third-party libraries
in iOS apps. 
We propose and implement a system with two-level profiles to detect
the existence of libraries and their versions (or range of versions.)
We solve the challenge of code duplication between libraries and implement
a prototype system. The evaluation result shows the accuracy of our system
and practical usage scenarios. 


%
%

\bibliographystyle{ACM-Reference-Format}
  \bibliography{paper}
\end{document}